\documentclass[12pt,pre,aps,superscriptaddress]{revtex4}
\usepackage{graphicx}
\newcommand{\beq}[2]{\begin{equation}#1\label{#2}\end{equation}}
\newcommand{\ceq}[1]{(\ref{#1})}

\newfont{\mbld}{cmbx10 scaled 800}
\newfont{\cab}{cmsy10 scaled 1200}
\newfont{\scab}{cmsy10 scaled 1000}
\newfont{\bcall}{cmbsy10 scaled 1200}

\begin{document}
\title{On a relation between Liouville field theory and a two
  component
scalar field theory passing through the random walk} 
\author{Franco Ferrari}
\email{ferrari@univ.szczecin.pl}
\affiliation{Institute of Physics and CASA*, University of Szczecin,
  ul. Wielkopolska 15, 70-451 Szczecin, Poland}
\affiliation{Laboratoire de Physique Mol\'eculaire et des Collisions,
Universit\`e Paul Verlaine-Metz, 1 bvd Arago, 57078 Metz, France}
\author{Jaros{\l}aw Paturej}\email{jpaturej@univ.szczecin.pl}
\affiliation{Institute of Physics and CASA*, University of Szczecin,
  ul. Wielkopolska 15, 70-451 Szczecin, Poland}
\affiliation{Laboratoire de Physique Mol\'eculaire et des Collisions,
Universit\`e Paul Verlaine-Metz, 1 bvd Arago, 57078 Metz, France}
%\author{Vakhtang G. Rostiashvili}\email{rostiash@mpip-mainz.mpg.de}
%\author{Thomas A. Vilgis} 
%\email{vilgis@mpip-mainz.mpg.de}
%\affiliation{Max Planck Institute for Polymer Research, 10
%  Ackermannweg, 55128 Mainz, Germany}

\begin{abstract}
In this work
it is proposed
a transformation which is useful
in order to
simplify non-polynomial
potentials  given in the form of an exponential.
As an application, it is shown that the quantum
Liouville field theory may be mapped into a field theory with a
polynomial interaction
between
two scalar fields and
a  massive vector field.

%In particular, it is shown that the grand canonical partition function
%of Brownian particles interacting with a massive vector field
%coincides with the partition function of Liouville field theory.
\end{abstract}
\maketitle
\section{Introduction}\label{sec:intro}
The Liouville field theory is actively studied both in physics and
mathematics 
\cite{Pol3,Pol4,N,LAG,Mu1,CD1,Ma1,Pol5,TA1,TA2,TZ1,Ma2,CMS,HJP}.
Several
approaches have been proposed for its quantization, see for example
Refs.~\cite{N,Tes1JLG}
for a list of references on this subject.
%\cite{CurTh,DH,DHF}. \cite{ZZ2,DO,T1,T3,GL},
%\cite{TA1,TA2,Ta3,Ta4,Mat,BBGMN,MT2,MT3},
%\cite{G,CGR1,CGR2,GS1,GS2,GR,T4}.

Motivated by the difficulties 
encountered in carrying out the quantization program of the Liouville model
posed by
the exponential interaction term present in its action, we show here
the equivalence between Liouville field theory
and a field theory with polynomial action
which describes the interaction of two scalar fields
with a massive vector field. The 
original degrees of freedom of the Liouville model are conveniently
mapped into the longitudinal component of the vector field, while
the spurious transverse components decouple after taking a suitable limit of
large coupling constant.
The appearance of the mass term of the vector field is related to the
kinetic energy of the Liouville field theory.
The two scalar fields are instead associated to constraints, so that they do
not introduce other, unwanted, degrees of freedom.
The advantage of the new theory derived in this way
is that it has a structure similar to
that of massive scalar electrodynamics and
it is thus much more tractable than Liouville field theory with
standard techniques, like for instance the perturbative
approach~\footnote{We would like to mention at 
  this point the fact 
  that a perturbative approach may be defined also for the
  quantum theory of Liouville, see
  \cite{jackiw1,jackiw2}.}.

At the heart of the mapping of the Liouville model into a polynomial
field theory there is a
transformation which allows to rewrite the exponential term typical of
Liouville field theory in the form of the generating functional of a
two component scalar field theory. The price to be paid
is the introduction of new degrees of freedom, 
however the scalar sector of the theory  is almost trivial because,
as already mentioned, its role is just to impose
constraints. 
Solvable theories in which  the number of degrees of freedom is zero
have been known for more than thirty years \cite{IBB}.
The technique of expressing complicated potentials in terms of
amplitudes of almost trivial field theories has been presented
in Ref.~\cite{franco2} in the case of topological
interactions in polymer physics. The advantages of this technique
in modeling the
entanglement of two or more polymers have been explained in
Refs.~\cite{francoN,francoann}. 

The mapping of Liouville field theory into the two component scalar
field
theory coupled to a vector field
is based in part on known results of the theory of Brownian
motion. In particular, it is applied the field theoretical
representation 
of the
grand canonical
partition function
of charged particles subjected to a random walk while immersed in a
magnetic field. 
This is not the first time that 
Liouville field theory has been related  to
statistical systems.
For example, in the works of
Refs.~\cite{Mu1,CD1}
the theory of Liouville appears either like a random
field or as a mean to express the normalization of the ground state
of a model of disordered conductors.
In \cite{Ma1, Mir1} 
it is associated to quantum
chaos in anyon systems. 
However, 
the main interest of this work is
not the 
connection between Liouville field theory and statistical
mechanics. The goal is instead to map the
Liouville field theory into a theory which is simpler in the sense
that its action is a polynomial in the fields, so that
field theoretical techniques like the perturbative approach can be
applied to it.  
The statistical theory of the random walk enters in our approach
 because it offers a nice way to express the Green function of fields
coupled to vector fields, a fact already noted in
Ref.~\cite{HalpernSiegel}. The application of these ideas to Liouville
field theory and the derivation of the equivalent model with polynomial
interactions constitutes the original part of the present work.

%In \cite{Ma1} the
%by various arguments, in particular by identifying the quantum
%Hamiltonian of anyons with the Weil--Petersson Laplacian \cite{Ma1}
%and by showing that 
%the equations of motion of anyons \cite{Ma1,Ja1}.
%Refs.~\cite{Mu1,CD1} 
%Liouville field theory has been
%interpreted as a random system. 
%In Refs.~
%the strict
%relationships between Liouville field theory and anyon systems has
%been proved, in particular in the case of static anyon solutions
%satisfying self-duality relations.
%However, this is the first time that 

This paper is organized as follows.
In Section II are presented the basic facts of the statistical
mechanics of Brownian motion which will be used later.
In Section III the partition function of a two component scalar field
theory interacting with 
a massive vector field is introduced using a path integral formulation.
In the limit in which the strength
of the interactions of the massive vector fields becomes infinite, their
transverse degrees of freedom decouple from the other fields and only
the longitudinal component survives.
In Section IV the two scalar fields are integrated out from the
partition function.
The result is a field theory in which there is an unique
scalar field with 
an exponential interaction term. The equivalence of this
field theory with the model of Liouville is proved in Section V.
Our conclusions are drawn in Section VI. Finally, in the Appendix a
procedure to 
integrate out the scalar fields is presented, which is alternative to
that of Section IV.
It is shown that the results of both procedures coincide.
%The crucial ingredient of our approach is a  Gaussian formula which
%allows to simplify the expression of potentials given in the
%exponential form as it happens in 
% Liouville field theory. 

\section{The Grand canonical partition function of Brownian particles in a
massive vector field}\label{sec:gracan}
Let  $t$ and $x^1,x^2$ be respectively the time and a set of
coordinates on the 
two dimensional plane $\mathbf R^2$.  A point in $\mathbf R^2$ is
denoted with the 
related  radius vector $\mathbf r=(x^1,x^2)$. 
The starting point of the discussion is the following differential
equation:
\beq{
\left[
\frac{\partial}{\partial t}-g\left(
\mathbf\nabla -b\mathbf A
\right)^2
\right]\Psi(t;\mathbf r,\mathbf r_0,\mathbf A)=\delta(t)\delta(\mathbf
r-\mathbf r_0)
}{langelike}
where $\mathbf A(\mathbf r)$ is a time independent vector field, whose
action will be specified later.
For
the moment, we note that the above equation is invariant under the
local transformations:
\begin{equation}
\Psi(t;\mathbf r,\mathbf r_0,\mathbf A)=e^{-b(\gamma(\mathbf r) -
  \gamma(\mathbf 
  r_0))}\Psi'(t;\mathbf r,\mathbf r_0,\mathbf A')\qquad\qquad
\mathbf A(\mathbf r)=\mathbf A'(\mathbf r)+\mathbf \nabla\gamma(\mathbf
  r)\label{loctra}
\end{equation}
in which $\gamma(\mathbf r)$ is an arbitrary function of $\mathbf r$.
The solution of Eq.~\ceq{langelike} may be expressed in terms of a
Feynman path integral:
\beq{
\Psi(t;\mathbf r,\mathbf r_0,\mathbf A)=\theta(t)\int_{\mathbf R(t)=\mathbf
  r\atop
\mathbf R(0)=\mathbf r_0}
{\cal D}\mathbf R(\sigma)\exp\left\{
-\int_0^t d\sigma\left[
\frac 1{4g}\dot\mathbf R^2+b\dot\mathbf R\cdot \mathbf A
\right]
\right\}
}{patint}
$\theta(t)$ being the Heaviside function: $\theta(t)=0$ if $t<0$ and
$\theta(t)=1$ if $t\ge0$. 
Eq.~\ceq{patint}
is related to the 
canonical partition function of a particle interacting with a vector
potential $\mathbf A$ and
performing a random walk in the plane.
%This partition function
%measures the probability that such a particle
%diffuses from the
%point $\mathbf r_0$ to the point $\mathbf r$ during the time $t$.
 Following Ref.~\cite{zinnjustinbook}, we introduce now the generating
functional of the correlation functions of the field $\Psi$:
\beq{
\Xi[J]=\left\langle
\exp\left[\int dtd\mathbf r J(t,\mathbf r)\Psi(t;\mathbf r,\mathbf
  r_0,\mathbf A)\right]
\right\rangle_{\mathbf A}
}{noiave}
%In Eq.~\ceq{noiave} it is understood that $\Psi(t;\mathbf r,\mathbf
%r_0)$ satisfies Eq.~\ceq{langelike} and thus depends on the field
%$\mathbf A$. 
In Eq.~\ceq{noiave} the average over the vector potential $\mathbf A$
is taken according to the
following prescription:
\beq{
\langle\cdots\rangle_{\mathbf A}=\int{\cal D}\mathbf A e^{\int
  d\mathbf r\left[
\alpha(\mathbf \nabla\times\mathbf A)^2+\frac{\mathbf A^2}2
\right] }\cdots
}{noigen} 
Putting
\beq{
J(t,\mathbf r)=\mu\delta(t-T)
}{jtwocurr}
we get from Eq.~\ceq{noiave}:
\beq{
\left.\Xi[J]\right|_{J=
\mu\delta(t-T)}\equiv\Xi[\mu]=\left\langle
\exp\left[
\mu\int d\mathbf r\Psi(T;\mathbf r,\mathbf r_0,\mathbf A)
\right]
\right\rangle_{\mathbf A}
}{gcparfun}
Expanding the exponential in the right hand side of Eq.~\ceq{gcparfun}
we have:
\beq{
\exp\left[
\mu\int d\mathbf r\Psi(T;\mathbf r,\mathbf r_0,\mathbf A)
\right]=\sum_{N=0}^\infty\frac1{N!}Z_N
(T;\mathbf r_0,\mathbf A)
\mu^N
}{forone}
where
\beq{
Z_N(T;\mathbf r_0,\mathbf A)=\left(
\int d\mathbf r\Psi(T;\mathbf r,\mathbf r_0,\mathbf A)
\right)^N
}{fortwo}
%$\Psi(T;\mathbf r,\mathbf
%r_0)$ as  evolution kernel, it is possible to conclude that 
The
quantity
$Z=\int d\mathbf r\Psi(T;\mathbf r,\mathbf r_0,\mathbf A)$ has the meaning of
the canonical partition function of
a particle diffusing 
from a fixed point $\mathbf r_0$ to any other
point in the plane during the  time $T$. 

From the above discussion it
turns out that $\Xi[\mu]$
can be interpreted as the grand canonical partition function of a
system of indistinguishable  
particles which perform a Brownian walk while they are interacting with
a massive vector field $\mathbf A$. The parameter $\mu$ plays the role
of the chemical potential. 
%Let us stress the fact that the sole reason
%for which the point $\mathbf r_0$ is kept fixed is that we wish to
%establish a relation between Eq.~\ceq{gcparfun}  and the partition
%function of the Liouville field theory. One could easily
%eliminate this constraint
%by introducing in Eq.~\ceq{noiave} also a sum over 
%all possible values of the initial point $\mathbf r_0$, but in this
%case the relation would be with a field theoretical model whose action
%is
%nonlocal.
\section{The massive vector field theory}\label{sec2}
Hereafter we will conform our notation to that
used in the case of Euclidean field theories in natural
units $\hbar=c=1$.
Points in $\mathbf R^2$ will be denoted with the symbols
$x,y,z,\ldots$, where $x=x^1,x^2$, $y=y^1,y^2$ 
etc.
Moreover, the spatial components of vectors will be labeled using
middle Greek indices $\mu,\nu,\ldots=1,2$.
The summation convention $u_\mu v^\mu=u_1v^1+u_2v^2$ will be used.
The volume measure $d\mathbf r$ is replaced by  $d^2 x$, the symbol
$(x-x_0)^2$ denotes the scalar product $(x-x_0)_\mu(x-x_0)^\mu$ and
similarly $\partial^2=\partial_\mu\partial^\mu$ is the laplacian and so on.

The main subject of this Section is a model of three dimensional
non-relativistic 
scalar fields 
$\psi_1(t,x),\psi_2(t, x)$ coupled
with a two dimensional 
massive vector field $A_\mu(x)$. 
The dimensions in natural units of these fields are respectively:
\beq{
[\psi_1]=0\qquad\qquad [\psi_2]=-3 \qquad\qquad [A_\mu]=-1
}{fiedim}
The  partition function of the model is given by:
\beq{
{\cal Z}_{\alpha,T}[J_1,J_2]=
\int {\cal D} A
{\cal D} \psi_1{\cal D} \psi_2
\exp{\left[-
S_\alpha(A,\psi_1,\psi_2)
\right]}
}{parfunzajojt}
where the action $S_\alpha(A,\psi_1,\psi_2)$ is:
\begin{eqnarray}
S_{\alpha,T}(A,\psi_1,\psi_2)
&=&
\int d^2x\left[
\alpha\frac{F_{\mu\nu}^2}4+\frac{A_\mu A^\mu}{2}
\right]
+
\int dtd^2x \left[
-iJ_1\psi_1+J_2\psi_2
\right]
\nonumber\\
&+&iT\int dtd^2x\left[
\psi_1\frac{\partial\psi_2}{\partial t}+g(\partial_\mu+bA_\mu)\psi_1
(\partial^\mu-bA^\mu)\psi_2
\right] 
\label{actsta}
\end{eqnarray}
and $F_{\mu\nu}=\partial_\mu A_\nu-\partial_\nu A_\mu$.
In  the above equation $J_1$ and $J_2$ represent  
external currents, while $\alpha$ and $T$ are real and positive
parameters. Both $T$ and the coupling constant $g$ have the dimension
of a length, while $\alpha$ has the dimension of a 
squared mass. The connection with the grand canonical partition
function of particles subjected to a Brownian motion is in the right
hand side of Eq.~\ceq{actsta}. For suitable choices of the currents
$J_1$ and $J_2$ the integration over the fields $\psi_1$ and $\psi_2$
reproduces exactly the grand canonical partition function of
Eq.~\ceq{gcparfun}.
%The massless version of the  theory described
%by the action \ceq{actsta} is:
%\begin{eqnarray}
%\!\!\!\!\!\! S_{\alpha,T}^{dil}(A,\psi_1,\psi_2)
%&=&
%\int d^2x\left[
%\alpha\frac{F_{\mu\nu}^2}4
%\right]
%+iT\int dtd^2x\left[
%\psi_1\frac{\partial\psi_2}{\partial t}+g(\partial_\mu+bA_\mu)\psi_1
%(\partial_\mu-bA_\mu)\psi_2
%\right]% \nonumber\\
%%&+&\int dtd^2x \left[
%%J_1\psi_1+J_2\psi_2
%%\right]
%\label{actstamasles}
%\end{eqnarray}
%Eq.~\ceq{actstamasles} does  not coincide with  the usual action of scalar
%electrodynamics with gauge group $U(1)$, 
%but rather it corresponds to a theory which is invariant under
%local rescaling of the fields:
%\beq{
%\psi_1\longrightarrow e^{-b\gamma}\psi_1\qquad\qquad
%\psi_2\longrightarrow e^{b\gamma}\psi_2\qquad\qquad
%A_\mu\longrightarrow A_\mu+\gamma
%}{diltra}
%with $\gamma=\gamma(x)$ being a function of the spatial coordinates
%$x^1,x^2$ only. Examples of field theories of this kind have been
%discussed in \cite{Fe1}.

Clearly, the free action of the massive vector fields
$
S^0_{mvf}(A)=\int d^2x\left[
\alpha\frac{F_{\mu\nu}^2}4+\frac{A_\mu A^\mu}2
\right]
$
corresponds to
the action of the vector field $\mathbf A(\mathbf r)$ in
Eq.~\ceq{noigen}. Moreover, in the action 
\ceq{actsta} the field $\psi_1$ plays the role of a Lagrange
multiplier which imposes the constraint:
\beq{
T\left[
\frac{\partial}{\partial t}
-g(\partial-bA)^2
\right]\psi_2=J_1
}{aconstr}
The above equation becomes equal to the differential equation
\ceq{langelike} upon making for the current $J_1$ the special choice
\beq{
J_1(t,x)=\frac Tg\delta(t)\delta(x-x_0)
}{specchocur}
and 
rescaling the field $\psi_2$ 
by the constant $g$, so that $\Psi=g\psi_2$.
As a consequence the partition function $Z_{\alpha,T}[J_1,J_2]$
defined by Eqs.~\ceq{parfunzajojt} and \ceq{actsta} is equivalent to
the generating functional of Eq.~\ceq{noiave}.

We are interested to study the 
limit in which 
both $\alpha$ 
and $T$ approach infinity. 
It is easy to check that the limit
$\alpha\longrightarrow +\infty$ imposes 
in the partition function \ceq{parfunzajojt}
the constraint:
\beq{
\epsilon^{\mu\nu}\partial_\mu A_\nu=0
}{constr}
where $\epsilon^{\mu\nu}$ is the totally antisymmetric tensor defined
according to the usual convention $\epsilon^{12}=1$.
As a matter of fact,
remembering that in two dimensions
$F_{\mu\nu}^2= 2F_{12}^2$, 
it is possible to apply to Eq.~\ceq{parfunzajojt} 
the following Gaussian identity:
\beq{
\exp\left[-\alpha\int d^2x \frac{F_{\mu\nu}^2}4\right]=
\int{\cal D}\lambda \exp\left[-\int d^2x
\left(i\lambda F_{12}+
\frac{\lambda^2}\alpha
\right)\right] 
}{gauide}
If $\alpha$ goes to infinity, in the right hand side of
Eq.~\ceq{gauide} 
the field $\lambda=\lambda(x)$ becomes a Lagrangian multiplier which
imposes exactly condition \ceq{constr}, because
$F_{12}=\epsilon^{\mu\nu} \partial_\mu A_\nu$.
Thus, when $\alpha$ goes to infinity only the longitudinal component
of the field $\mathbf A$ survives due to the identity \ceq{constr},
which eliminates the transverse components.
The meaning of the limit
$T\longrightarrow+\infty$ 
will become
clear later,  
as we will introduce a second Gaussian identity resulting after
the integration of the
fields $\psi_1$ and $\psi_2$.

\section{The limit $\alpha=+\infty$}\label{sec:class}
In this Section we consider the model described 
in the previous Section
in the limit $\alpha=+\infty$. As it has been discussed earlier, in
this case the condition \ceq{constr} 
forces the field $A_\mu$ to have just the longitudinal component,
i. e.
$A_\mu=\partial_\mu\varphi$.
As a consequence, the partition function
\ceq{parfunzajojt} and the action \ceq{actsta} become respectively:
\beq{
{\cal Z}_{\infty,T}[J_1,J_2]={\cal C}
\int {\cal D} \varphi
{\cal D} \psi_1{\cal D} \psi_2
\exp{\left[-
S_{\infty,T}(\varphi,\psi_1,\psi_2)
\right]}
}{parfunainf}
and
\begin{eqnarray}
S_{\infty,T}(\varphi,\psi_1,\psi_2)
&=&
\int d^2x\left[
\frac{\partial_\mu\varphi \partial^\mu\varphi}{2}
\right]+\int dtd^2x \left[
-iJ_1\psi_1+J_2\psi_2
\right]
\nonumber\\
&+&
iT\int dtd^2x\left[
\psi_1\frac{\partial\psi_2}{\partial
  t}+g(\partial_\mu+b\partial_\mu\varphi)\psi_1 
(\partial^\mu-b\partial^\mu\varphi)\psi_2
\right] 
\label{actstainf}
\end{eqnarray}
In Eq.~\ceq{parfunainf} ${\cal C}=\det\partial_\mu$ is an irrelevant
constant which appears because of the change of measure ${\cal
  D}(\partial_\mu\varphi)\longrightarrow {\cal D}\varphi$
and will be omitted in the following. 

At this point we perform in the partition function \ceq{parfunainf}
the field redefinitions: 
$
\psi_1=e^{-b\varphi}\psi_1'$ and
$\psi_2=e^{b\varphi}\psi_2'$.
As a result we obtain:
\beq{
{\cal Z}_{\infty,T}[J_1,J_2]=
\int {\cal D} \varphi
{\cal D} \psi_1'{\cal D} \psi_2'
\exp{\left[-
S_{\infty,T}(\varphi,\psi_1',\psi_2')
\right]}
}{parfunfiered}
with
\begin{eqnarray}
S_{\infty,T}(\varphi,\psi_1',\psi_2')
&=&
\int d^2x\left[
\frac{\partial_\mu\varphi \partial^\mu\varphi}{2}
\right]+\int dtd^2x \left[
-iJ_1e^{-b\varphi}\psi_1'+J_2e^{b\varphi}\psi_2'
\right]
\nonumber\\
&+&
iT\int dtd^2x\left[
\psi_1'\frac{\partial\psi_2'}{\partial
  t}+g\partial_\mu\psi_1'
\partial^\mu\psi_2'
\right] 
\label{actstafiered}
\end{eqnarray}
The action
 \ceq{actstafiered} is Gaussian in the fields
$\psi_1'$ and
$\psi_2'$ and
it is thus possible
to eliminate these fields
 with the help of a simple Gaussian integration in the partition function
\ceq{parfunfiered}. 
To this purpose, we have to compute the path integral:
\begin{eqnarray}
{\cal Z}_T[\varphi, J_1, J_2]&=&\nonumber\\
&&\!\!\!\!\!\!\!\!\!\!\!\!\!\!\!\!\!\!\!\!
\!\!\!\!\!\!\!\!\!\!\!\!\!\!\!\!\!\!\!\!\!\!\!\!\!
\int
{\cal D} \psi_1'{\cal D} \psi_2'
\exp\left\{-
\int dtd^2x\left[
iT\left(\psi_1'\frac{\partial\psi_2'}{\partial
  t}+g\partial_\mu\psi_1'
\partial^\mu\psi_2'\right)-ie^{-b\varphi} J_1\psi_1'+
e^{b\varphi}J_2\psi_2' 
\right] 
\right\}
\label{auxpatint}
\end{eqnarray}
Here with the symbol ${\cal Z}_T[\varphi,J_1, J_2]$ we
have denoted the part of the partition function ${\cal
  Z}_{\infty,T}[J_1,J_2]$ which contains only the fields $\psi_1'$ and
$\psi_2'$, i.~e.:
\beq
{
{\cal
  Z}_{\infty,T}[J_1,J_2]=\int{\cal D}\varphi
e^{-\int d^2x\frac 12\partial_\mu \varphi\partial^\mu\varphi}
{\cal Z}_T[\varphi, J_1, J_2]
}{didiii}
%where now the currents  $\tilde J_1$ and $\tilde J_2$ are given by:
%\beq{
%\tilde J_1=J_1e^{-b\varphi}\qquad\mbox{and}\qquad
%\tilde J_2=J_2e^{b\varphi}
%}{newcur}

In the path integral \ceq{didiii} the field $\psi_1'$ plays the role
of a Lagrange multiplier. After integrating it out, the generating
functional ${\cal Z}_T[\varphi, J_1, J_2]$ becomes:
\beq{
{\cal Z}_T[\varphi, J_1, J_2]=\int{\cal D}\psi_2'\delta\left(
T\frac{\partial \psi'_2}{\partial t}-g\triangle\psi_2' -e^{-b\varphi}J_1
\right)\exp\left\{-\int dt d^2x e^{b\varphi} J_2\psi_2'\right\}
}{tralala}
The $\delta-$function appearing in the above equation forces the field
$\psi_2'$ to satisfy the equation:
\beq{
T\frac{
\partial \psi_2'}{\partial t}
-g\triangle\psi_2'=e^{-b\varphi }J_1
}{eqggg}
At this point we make for the current $J_1$ the special choice:
\beq{
J_1(t,x)=\frac Tg \delta (t)\delta(x-x_0)
}{jonesc} 
The solution of Eq.~\ceq{eqggg} corresponding to this choice is:
\beq{
\psi_2'(t,x)=\theta(t)\frac{T}{
4\pi g^2 t}e^{-\frac{
(x-x_0)^2
}{4gt}}e^{-b\varphi(x_0)}
}{lsss}
Thus, when integrating over the field $\psi_2'$
in Eq.\ceq{tralala}, we arrive at the following expression of the
generating  functional ${\cal Z}_T[\varphi, J_1, J_2]$:
\beq{
{\cal Z}_T[\varphi, J_1, J_2]=
\exp\left[-
\int dt d^2x J_2(t,x)\frac{
\theta(t)T
}{4\pi g^2t}e^{b(\varphi(x)-\varphi(x_0))}
e^{-\frac{(x-x_0)^2}{4gt}}
\right]
}{ghgj}
It is still possible to use the freedom to choose the second current $J_2$:
\beq{J_2(t,x)=T\delta(t-T)}{jtwosc}
In this way we obtain the final form of the functional 
${\cal Z}_T[\varphi, J_1, J_2]$:
\beq{
{\cal Z}_T[\varphi, J_1, J_2]=\theta(T)\exp\left[-
\int d^2x \frac{
e^{b(\varphi(x)-\varphi(x_0))}
}{4\pi g^2}
e^{-\frac{(x-x_0)^2}{4gT}}
\right]
}{finintfun}
More generally, what we have proved here is the following 
identity:
\begin{eqnarray}
&&\int{\cal D}\psi_1{\cal D}\psi_2\exp\left\{
-\int dt d^2x\left[iT\left(
\psi_1\frac{\partial \psi_1}{\partial
  t}+g(\partial_\mu+b\partial_\mu\varphi )
(\partial^\mu-b\partial^\mu\varphi )
\psi_2
\right)+J_2\psi_2-iJ_1\psi_1
\right]
\right\}\nonumber\\
&=&\exp\left\{
-\int dt d^2x\int dt' d^2x'\left[
e^{b(\varphi(x)-\varphi(x'))}\frac{\theta(t-t')}{4\pi gT(t-t')}
e^{\frac{ (x-x')^2}{4g(t-t')}}J_1(t',x')J_2(t,x)
\right]
\right\}
\label{finintfuntwo}
\end{eqnarray}
of which Eq.~\ceq{finintfun} is a particular case.
Putting the result of Eq.~\ceq{finintfun} back in Eq.~\ceq{didiii}, we find:
\beq{
{\cal Z}_{\infty,T}[J_1,J_2]=
\int{\cal D}\varphi\exp\left\{
-\int d^2x\left[
\frac{(\partial_\mu\varphi)^2}{2}+e^{b(\varphi(x)-\varphi(x_0))}
\frac{\theta(T)}{4\pi
 g^2}
\exp\left( 
-\frac{(x-x_0)^2}{4gT}\right)
\right]
\right\}
}{parfunqualio}

%After a few calculations and apart from an irrelevant constant, one finds:
%\beq{{\cal Z}[\tilde J_1,\tilde J_2]=
%\exp\left\{-
%\int_{-\infty}^{+\infty}dt\int_{-\infty}^{+\infty}dt'
%\int d^2x\int d^2y \tilde J_2(t,x){\cal G}(x-y;t-t')\tilde J_1(t',y)
%\right\}
%}{reszcalint}
%The propagator ${\cal G}(x-y;t-t')=\langle0|
%\psi_1'(x,t)\psi_2'(t',y)
%|0\rangle$ of the $\psi'-$fields
%  satisfies the differential equation:
%\beq{
%T\left(
%\frac{\partial}{\partial t}-g\Delta 
%\right){\cal G}(x-y;t-t')=\delta(t-t')\delta(x-y)
%}{equpro}
%The solution of Eq.~\ceq{equpro} is the Green function:
%\beq{{\cal G}(x-y;t-t')=\frac{\theta(t-t')}{4\pi
% gT(t-t')}\exp\left\{ 
%-\frac{(x-y)^2}{4g(t-t')}
%\right\}
%}{grefun}
%We are now ready to perform the Gaussian integration over the fields
%$\psi_1',\psi_2'$ in Eq.{auxpatint}.
\section{The limit $T\longrightarrow \infty$ and  Liouville field
  theory}\label{sec:finlim} 
%So far, the currents $J_1,J_2$ appearing in
%Eq.~\ceq{newcur} have been arbitrary. In this Section we will make the
%special choice:
%\beq{
%J_1(t',y)=\frac Tg\delta(t')\delta(y-x_0)
%\qquad\qquad J_2(t,x)=T\delta(t-T)
%}{specho}
%where $x_0$ is any fixed point in the two dimensional space.
%The first of Eqs.~\ceq{specho} 
%coincides with Eq.~\ceq{specchocur}, while the second one is exactly
%equal
%to Eq.~\ceq{jtwocurr}
%provided $\mu=T$.
%With this particular choice the partition function \ceq{parfunfiered}
%becomes:

In this Section the limit $T\longrightarrow +\infty$ is taken
in the partition function
\ceq{parfunqualio}. In this limit it
is possible to put $\theta(T)=1$ since $T>0$. Using the fact that
$\lim_{T\to+\infty}
\exp\left\{ 
-\frac{(x-x_0)^2}{4gT}\right\}=1
$
%As a consequence, taking the limit $T\longrightarrow+\infty$ in
%Eq.~\ceq{parfunqualio}, 
we arrive in this way at the partition function:
\beq{
{\cal Z}_{\infty,\infty}[J_1,J_2]=
\int{\cal D}\varphi\exp\left\{
-\int d^2x\left[
\frac{(\partial_\mu\varphi)^2}{2}+
\frac{1}{4\pi
 g^2}e^{b(\varphi(x)-\varphi(x_0))}
\right]
\right\}
}{parfunlinf}
Finally, performing the shift of fields $\varphi'(x)=\varphi(x)-\varphi(x_0)$,
one finds the partition function of the Liouville model with the
additional condition 
\beq{
\varphi'(x_0)=0
}{addconstr}
It is easy to check that in the limits $b=0$ and
$g\longrightarrow+\infty$
the partition function in Eq.~\ceq{parfunzajojt} becomes trivial, as
it is expected from the equivalence with the above Liouville field theory.
\section{Conclusions}
In this work the Liouville field theory has been mapped into the two
component
scalar
field theory interacting with 
a massive vector field defined by Eqs.~\ceq{parfunzajojt}  
and \ceq{actsta}. The two theories have been proved
to be equivalent. The massive vector field model after the limit
$\alpha\longrightarrow +\infty$ may be considered in
some sense as a sort of BF model \cite{btrb} where just the
longitudinal modes propagate, since 
the gauge invariance is broken
by the mass-term~\footnote{We thank the anonymous referee of {\it
    Phys. Lett. B} for this important comment and for other helpful hints.}. 
%First,
%the generating functional of Eq.~\ceq{noiave} has been identified with
%the grand canonical partition function of a system of identical
%particles given by Eq.~\ceq{gcparfun}. The same generating functional
%has been
%rewritten
%as
%the partition function of scalar fields coupled to massive
%vector fields given in
%Eqs.~\ceq{parfunzajojt} 
%and \ceq{actsta}.
%In a second step it has been shown that, in the limit
%$\alpha,L\longrightarrow+\infty$ the partition function
%\ceq{parfunzajojt} coincides with that of a Liouville model.
The most important
ingredient in our procedure is the Gaussian formula
\ceq{finintfuntwo}. This formula 
 allows to express the exponential interaction
term in the Liouville action  in the form of
  the generating functional
of two scalar fields in three dimensions. 
Eq.~\ceq{finintfuntwo} has been obtained after the field rescaling
$
\psi_1=e^{-b\varphi}\psi_1'$ and
$\psi_2=e^{b\varphi}\psi_2'$. This procedure may in principle alter the
functional integral measure, possibly spoiling our result. For this
reason, in 
the Appendix Eq.~\ceq{finintfuntwo} has been re-derived using an
alternative method, 
which does not involve the rescaling of the fields. 

The model of  Eqs.~\ceq{parfunzajojt}
and \ceq{actsta} has polynomial interactions and can be treated by
standard field theoretical 
techniques. If one uses for instance
the method of 
Ref.~\cite{Fer1}, the massive vector
field $\mathbf A$ can be eliminated, leaving as a result a non-local
and multi-component scalar field theory. Despite the non-locality,
theories of this kind may be investigated with the help of
approximations like RPA or Hartree--Fock, or withing the techniques of
strong coupling \cite{kleinert}. 
One limitation of our
procedure is that 
the Liouville field $\varphi$ is  mapped into the longitudinal
component of the vector field $\mathbf A$. 
For this reason, the potential obtained in
Eq.~\ceq{parfunlinf} depends on the difference of fields
$\varphi(x)-\varphi(x_0)$. For the same reason, 
it is only possible to compute the correlation
functions of differences of Liouville fields. Alternatively, one may
require that the field  
 $\varphi$ satisfies the condition
\ceq{addconstr} at an arbitrarily chosen point
$x_0$ as it has been done in this work.

Let us note that the derivation of Eq.~\ceq{finintfuntwo}, as well as the
whole procedure used in order to obtain the theory of Liouville from
the model of Eqs.~\ceq{parfunzajojt}
is not just a sequence of formal passages. It is
actually based on the way in which in statistical
mechanics one passes from the canonical partition function to the grand 
canonical partition function, as it was briefly explained in
Section~\ref{sec:gracan}. Additionally, the grand canonical partition
function has been written with the help of scalar fields, following
 path integral techniques used in many-body physics \cite{negele}.
For this reason, our approach is not limited   to the Euclidean two
dimensional space,
but it may be extended to any
 $n-$dimensional manifold on which the solution of
Eq.~\ceq{langelike}, expressing
the Green function of a Brownian
particle immersed in an external magnetic field, is known.
Of course, the Liouville action will be no
longer renormalizable if $n>2$. This is consistent
with the fact that, within our 
procedure, the theory of Liouville is
mapped to a massive vector field theory, which is also nonrenormalizable
in dimensions higher than two.
In extending our approach
to general manifolds,
one should be aware that 
the Gaussian formula \ceq{finintfuntwo} is
modified by the geometry and/or by the presence of zero modes.
Already in the case of an $n-$dimensional Euclidean space,
the Green function
$\Psi(t;\mathbf r,\mathbf r_0,\mathbf A)$ satisfying
Eq.~\ceq{langelike}
has a behavior with respect to the time $t$
which depends on $n$. Therefore, if we wish to obtain
an analogue of the Liouville partition function 
in $n-$dimensions after performing the large time limit
$T\longrightarrow +\infty$ as it has been done in
Section~\ref{sec:finlim}, we need to 
change the constant coefficients appearing in the currents $J_1,J_2$.
On a general manifold the situation is much more complicated.
It is easy to adjust the left hand side of
Eq.~\ceq{finintfuntwo}, which is the partition function of a field
theory, to include a background metric. The same is not true however for the
right hand side, which is the solution of a differential equation
whose closed form is known just in the case of a few non-flat geometries. 
An additional difficulty can arise on spatial manifolds
which admit
non-trivial classical solutions of Eq.~\ceq{constr} consisting
of harmonic zero modes. On compact Riemann surfaces the explicit form
of these non-trivial classical solutions may be found in
Refs.~\cite{ferrari1,ferrari2}.
The problem is that the
limit $\alpha\longrightarrow +\infty$ does not project out the
harmonic zero modes from the action \ceq{actstainf}, so that one
should eliminate them manually. In gauge field theories the harmonic
zero modes may be regarded as gauge degrees of freedom and gauged away
using a
method based on BRST techniques
proposed by Polyakov \cite{polyakov} and further developed by
the authors of 
Refs.~\cite{amati1,gtmb}. In the present context, however, this strategy
cannot be applied. As a matter of fact, the gauge symmetry
\ceq{loctra} is explicitly broken by the mass term of the
gauge fields and by the insertion of the external currents
$J_1,J_2$, which are needed in order to obtain the theory of
Liouville in its final form. If we attempt to
gauge away the harmonic zero
modes with a transformation similar to the field rescaling which 
has been used in
order to obtain Eq.~\ceq{parfunfiered}, they will remain in the action
due to the
current terms.
%Even in the absence of non-trivial solutions of Eq.~\ceq{constr},
%on a manifold 
%the behavior of the Green function solving
% Eq.~\ceq{langelike} at large times could be different from that
%of the flat Green function satisfying Eq.~\ceq{langelike}.
%It is thus not granted that
%the large time limit $T\longrightarrow +\infty$ 
%performed in Section~\ref{sec:finlim}
%in order to recover the Liouville partition function
%will give the same result on a general manifold.

In conclusion, the extension of the present approach to nontrivial
manifolds is a difficult task, in particular because the explicit
expressions of the canonical and of the grand canonical partition
functions are not known in the case of a Brownian particle immersed
both in a magnetic field and in a gravitational field with an
arbitrary metric. As a
consequence, 
it is not possible
to write down an analogue of the fundamental identity
\ceq{finintfuntwo} on general manifolds. What it is however feasible, is 
%To conclude, we hope that it is  possible
the generalization of the Gaussian formula
\ceq{finintfuntwo} to include other theories than the Liouville model.
Work is in progress in that direction.
%for instance the Sine--Gordon model.  
\begin{appendix}
\section{An alternative proof of Eq.~\ceq{parfunqualio}}
\begin{figure}[h]
\includegraphics[width=.33\textwidth]{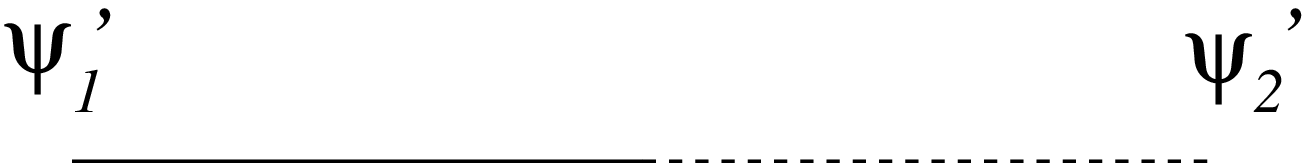}\hspace{2cm}
\includegraphics[width=.33\textwidth]{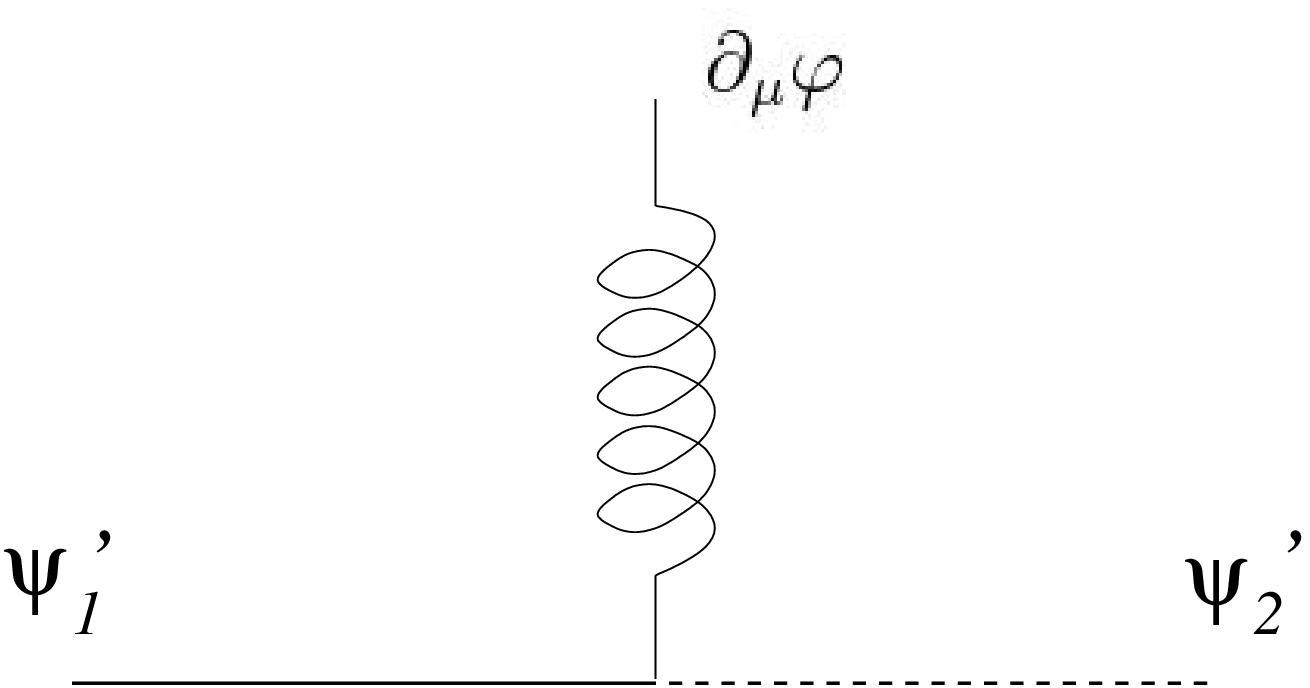}\\[1cm]
\includegraphics[width=.33\textwidth]{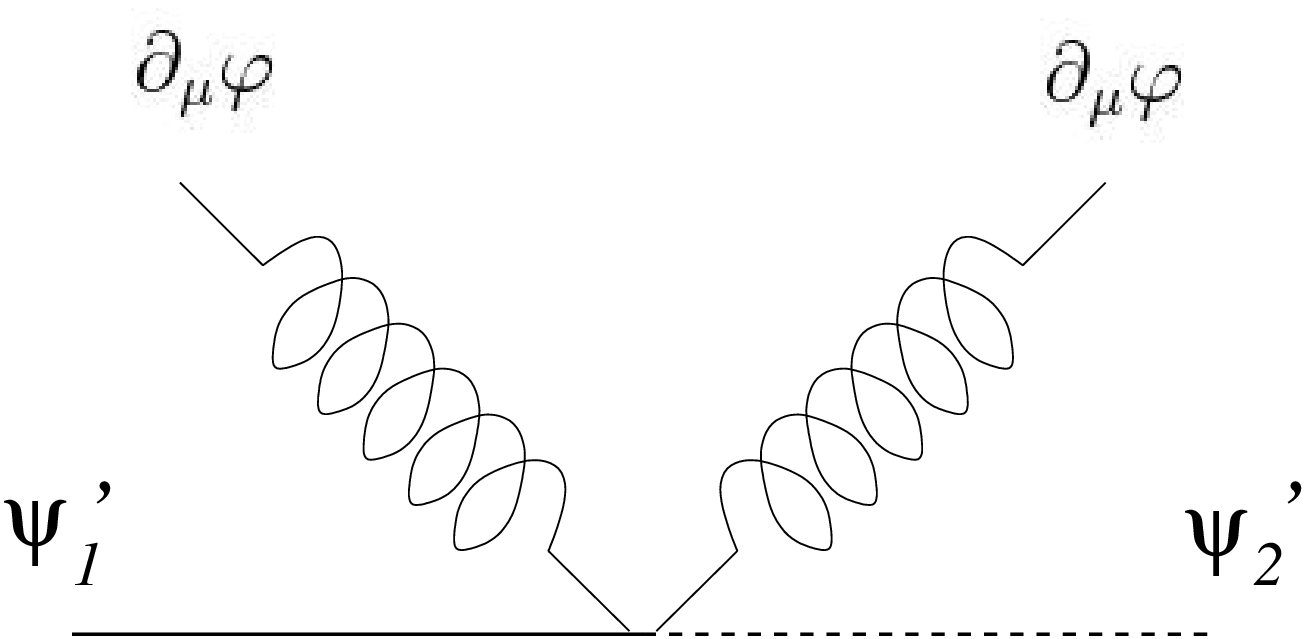}
\caption{Feynman diagrams of the  $\psi'_1,\psi_2'$
  action appearing in Eq.~\ceq{detkappa}. Full lines and dashed lines
  propagate 
  the fields $\psi_1'$ and $\psi_2'$ respectively, while wavy lines
  correspond to the insertion of the external field $\varphi$. }  
\label{feyruls}
\end{figure}

\begin{figure}[htbp]
\includegraphics[width=.99\textwidth]{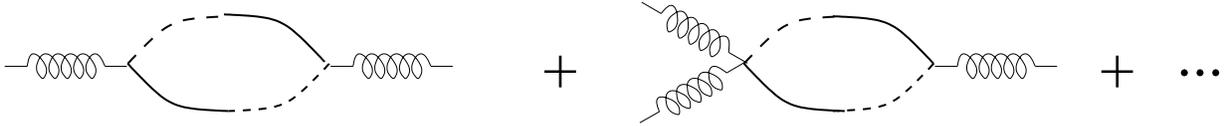}
\caption{Loop expansion of $\det K$.}
\label{loops}
\end{figure}

In this Appendix we provide an alternative method to perform the 
integration over the fields $\psi_1,\psi_2$ in the partition function
$Z_{\infty,T}[J_1,J_2]$ 
of Eq.~\ceq{parfunainf}. To this purpose,
we isolate from the expression of $Z_{\infty,T}[J_1,J_2]$ only
the part in which the fields $\psi_1,\psi_2$ are involved:
\begin{eqnarray}
Z_\psi&=&\int{\cal D}\psi_1{\cal D}\psi_2 \exp\left\{
-iT\int dt d^2x\left[
\psi_1\frac{\partial\psi_2}{\partial t}+
gD_{+\mu}\psi_1D_{-}^\mu\psi_2
\right]
\right\}\nonumber\\
&\times&\exp\int dt d^2x \left[
iJ_1\psi_1-J_2\psi_2
\right]\label{initial}
\end{eqnarray}
with
\beq{
D_{+\mu}=\partial_\mu+b\partial_\mu\varphi
\qquad\qquad
D_{-}^\mu=\partial^\mu-b\partial^\mu\varphi
}{covdevdef}
Since the action in \ceq{initial} is at most quadratic in the fields
$\psi_1,\psi_2$, it is convenient to perform the shift of variables:
\begin{eqnarray}
\psi_1(t,x)&=&\psi_1'(t,x)+i\int
dt^{\prime\prime}d^2x^{\prime\prime}G(t^{\prime\prime}-t;
x^{\prime\prime}-x;\varphi) 
J_2(t^{\prime\prime},x^{\prime\prime})\label{subone}  \\
\psi_2(t,x)&=&\psi_2'(t,x)+\int dt' d^2x'
G(t-t';x-x';\varphi)J_1(t',x')\label{subtwo} 
\end{eqnarray}
where
\beq{
G(t-t';x-x';\varphi)=\langle
\psi_1(t',x')\psi_2(t,x)
\rangle
}{fefffff}
is the two point function of the fields $\psi_1,\psi_2$:
\beq{
G(t-t';x-x';\varphi)=e^{b(\varphi(x)-\varphi(x'))}\frac{
\theta(t-t')
}{4\pi g T(t-t')} e^{-\frac{
(x-x')^2
}{4g(t-t')}}
}{twopfpp}
After the substitution (\ref{subone}--\ref{subtwo}), the partition
function $Z_\psi$ becomes:
\beq{
Z_\psi=\det K e^{
-\int d^2x\frac{\theta(T)}{4\pi g^2}e^{-\frac{(x-x_0)^2}{4gT}}
e^{b(\varphi(x)-\varphi(x_0))}
}
}{zpsifin}
with
\beq{
\det K=\det\left[
-iT\left(
\frac\partial{\partial t}-gD_{-\mu}D_-^{\mu}
\right)
\right]^{-1}
}{detedefi}
In order to obtain Eq.~\ceq{zpsifin} we have used the special values
of the currents $J_1,J_2$ given in Eqs.~\ceq{jonesc} and \ceq{jtwosc}
respectively.
Substituting the above result for $Z_\psi$ in the partition function
$Z_{\infty,T}[J_1,J_2]$ we find:
\begin{eqnarray}
Z_{\infty,T}[J_1,J_2]&=&\int{\cal D}\varphi 
e^{-\int d^2x
\frac{(\partial_\mu \varphi)^2}{2}}
\det K\nonumber\\
&\times& \exp\left\{
-\int d^2x\frac{\theta(T)}{4\pi g^2}e^{-\frac{
(x-x_0)^2
}{4g T}} e^{b(\varphi(x)-\varphi(x_0))}
\right\}\label{zinftnew}
\end{eqnarray}
Apart from the presence of the determinant of the differential
operator $K$, the expression of the partition function
$Z_{\infty,T}[J_1,J_2]$ is equal to that reported in
  Eq.~\ceq{parfunqualio}.
In principle, this determinant is dependent on the field $\varphi$ and
may thus lead to a theory which is not that of Liouville.
To show that this is not true because $\det K$ is trivial, we 
follow an analogous proof provided in the case of non-relativistic scalar
electrodynamics in Ref.~\cite{HalpernSiegel}. First of all, we
write the determinant
as a path integral in the fields $\psi_1',\psi_2'$:
\begin{eqnarray}
\det K&=& \int{\cal D}\psi'_1{\cal D}\psi'_2
\exp\left\{
iT\int dt d^2x\left[ \psi_1'\left(
\frac\partial{\partial t}-g\partial_\mu^2
\right)\psi_2'\right]\right\}\nonumber\\
&\times&\exp\left\{
iT\int dt d^2x\left[
gb\partial_\mu \varphi(
\psi'_1\partial^\mu\psi_2'+\partial_\mu\psi_1'\psi_2'
)-gb^2(\partial_\mu\varphi)^2\psi_1'\psi_2'
\right]
\right\}
\label{detkappa}
\end{eqnarray}
Here we have separated the contributions appearing in the action
according to the different powers of the coupling constant $b$.
In evaluating $\det K$, the field $\varphi$ may be regarded as an
external field.
The action appearing in the exponent of Eq.~\ceq{detkappa} produces
the Feynman rules shown in Fig.~\ref{feyruls}.
It is easy to realize that the Feynman diagrams which contribute to
$\det K$ are loops with $n$ external legs.
The external legs correspond to the
insertion of the external
fields $\varphi$, see Fig.~\ref{loops}.
Moreover, the loops consist
of dashed internal lines  alternated by full internal
lines. This is due to the fact that the free propagator
\beq{
\langle\psi_1'(t',x')\psi_2'(t,x)\rangle=\frac{\theta(t-t')}{4\pi g
  T(t-t')}
e^{-\frac{(x-x')^2}{4g(t-t')}}
}{proppsionetwo}
allows only the contractions of  $\psi_1'$ fields with 
$\psi_2'$ fields.  Most important, the propagator \ceq{proppsionetwo}
is multiplied by the
Heaviside
function $\theta(t-t')$. It is thus easy to convince oneself that the loops
which contribute to $\det K$ vanish identically because, after the
contractions of the fields $\psi_1'$ with the fields $\psi_2'$ are
performed, there is no common interval of time in which all the resulting
$\theta-$functions could be simultaneously different from zero. As a
consequence,
$\det K=1$ and the expression of the partition function
$Z_{\infty,T}[J_1,J_2]$ 
given in Eq.~\ceq{zinftnew} coincides with that of
Eq.~\ceq{parfunqualio}, ending our proof.
\end{appendix}
\begin{acknowledgments}
This work has been financed in part with funds allocated for the years
2006-2008 by the Polish Ministry of Science and University
(Ministerstwo Nauki i Szkolnictwa Wy\.zszego) in the frame of the
scientific project "Applications of supersymmetric and topological
field theories to the description of realistic polymer systems",
Scientific Grant no.  N202 156 31/2933.  
The authors would also like to thank Tomasz Wydro and  the LPMC, Universit\'e
Paul Verlaine-Metz, where the last stage of this work was performed,
for the nice hospitality. 

Last but not last F. Ferrari is indebted to C. Mudry 
and to M. Matone
for pointing out the interesting references
\cite{Mu1,CD1, Ma1}.

\end{acknowledgments}

\end{document}